# Graph for Science: From API based Programming to Graph Engine based Programming for HPC

Yu Zhang[1], Zixiao Wang[1], Jin Zhao[1], Yuluo Guo[1], Hui Yu[1], Zhiying Huang[1], Xuanhua Shi[1], Xiaofei Liao[1]

[1] National Engineering Research Center for Big Data Technology and System, Services Computing Technology and System Lab, Cluster and Grid Computing Lab, School of Computer Science and Technology, Huazhong University of Science and Technology, Wuhan, China

## ABSTRACT

Modern scientific applications predominantly run on large-scale computing platforms, necessitating collaboration between scientific domain experts and high-performance computing (HPC) experts. While domain experts are often skilled in customizing domain-specific scientific computing routines, which often involves various matrix computations, HPC experts are essential for achieving efficient execution of these computations on large-scale platforms. This process often involves utilizing complex parallel computing libraries tailored to specific matrix computation scenarios. However, the intricate programming procedure and the need for deep understanding in both application domains and HPC poses significant challenges to the widespread adoption of scientific computing. In this research, we observe that matrix computations can be transformed into equivalent graph representations, and that by utilizing graph processing engines, HPC experts can be freed from the burden of implementing efficient scientific computations. Based on this observation, we introduce a graph engine-based scientific computing (Graph for Science) paradigm, which provides a unified graph programming interface, enabling domain experts to promptly implement various types of matrix computations. The proposed paradigm leverages the underlying graph processing engine to achieve efficient execution, eliminating the needs for HPC expertise in programming large-scale scientific applications. We evaluate the performance of the developed graph compute engine for three typical scientific computing routines. Our results demonstrate that the graph engine-based scientific computing paradigm achieves performance comparable to the best-performing implementations based on existing parallel computing libraries and bespoke implementations. Importantly, the paradigm greatly simplifies the development of scientific computations on large-scale platforms, reducing the programming difficulty for scientists and facilitating broader adoption of scientific computing. The datasets used in the experiments as well as evaluation codes have been made available for download on GitHub at https://github.com/CGCL-codes/G4S.

## 1 INTRODUCTION

Matrix operations [3] represent the dominant cost of many scientific application domains [4]-[6], because these routines typically need to solve numerical integral equations, differential equations, etc. For example, the time for matrix multiplication and matrix addition accounts for more than 90% of the total execution time of LINPACK [5]. In order to cater to the specific needs of different scientific domains and deliver the maximum performance, domain experts often have to customize the computation of matrix operations in their domains. However, supporting high-performance matrix operations, which typically run on large-scale computing platforms [1], [2], demands extensive and continuous efforts from HPC specialists. As new hardware processing units (e.g., GPU and FPGA) emerge, more and more parallel computing libraries have been developed, such as MPI library [21], TensorFlow [22], cuBLAS [23], cuSPARSE [24], Intel MKL [25], ROCm [26], LAPACK [27], OpenBLAS [28], and numpy [29]. These libraries have been implemented into the existing library-based programming paradigm (e.g., MPI, cuBLAS) for scientific applications, providing respective APIs. On top of these libraries, many commercial softwares [30], [31], e.g., Ansys [32], Abaqus [33], COMSOL [34], Fluent [35], and STAR-CCM+ [36], have been developed for different scientific domains, such as molecular dynamics [7], fluid dynamics [8], electromagnetic [9], geodynamics [10], and chemical kinetics [11]. Although these software have achieved great successes in delivering high performance, this library-based programming paradigm for scientific computing still poses a great challenge.

Specifically, the input matrices of the matrix operations in different application domains may i) embody different characteristics, and ii) be performed on different hardware platforms. Studies show that there can be a more than 100 times performance gap between optimal and arbitrary implementations of matrix operations on large-scale computing platforms [37]. To achieve the optimized performance, the users have to heavily rely on their experience to i) select the use of the library that can best realize the characteristics of the matrix operations in question, ii) calls the intricate APIs provided by the library, and iii) on a case-by-case basis, optimize the implementation of scientific computing routine on large-scale computing platform with heterogeneous processing units [12]. The case-by-case optimization includes ensuring load balance and reducing communication cost, etc. All these necessitate deep involvement from experienced HPC experts, which hinders the rapid rollout of scientific applications. Consequently, developing a scientific computing routine takes a long time, and it may also require several years of continuous optimization to maintain high performance [38], [39].

In this paper, we observe that matrix operations can be viewed from the graph perspective and transformed into equivalent graph operations. Through the transformation, various matrix operations can be transformed to the graph operations that bear the unified interface. As the result, the users only need to utilize two unified programming interfaces: *Gather*() and *Apply*(), which are supported by any existing graph engine [13]-[17], to implement the matrix operations in various scientific computing routines.

Following on the observation, we develop a new programming paradigm, which we call G4S (Graph for Science), for scientific applications. G4S is a graph-based paradigm that simplifies the programming of high-performance scientific computing routines. G4S is fundamentally different from the traditional library-based programming paradigm, as it eliminates the heavy reliance on HPC experts and significantly shortens the development time. Moreover, the programming-friendliness of G4S does not sacrifice the performance of matrix operations. The data relationships inherent in the input matrices of different types of matrix operations (e.g., sparse, dense, symmetric, triangular, packed, banded and Hermitian matrix multiplications) are naturally exposed during their transformations to graphs, and therefore can be exploited automatically by the existing graph engines [40]-[46] to achieve optimal executions on large-scale computing platforms.

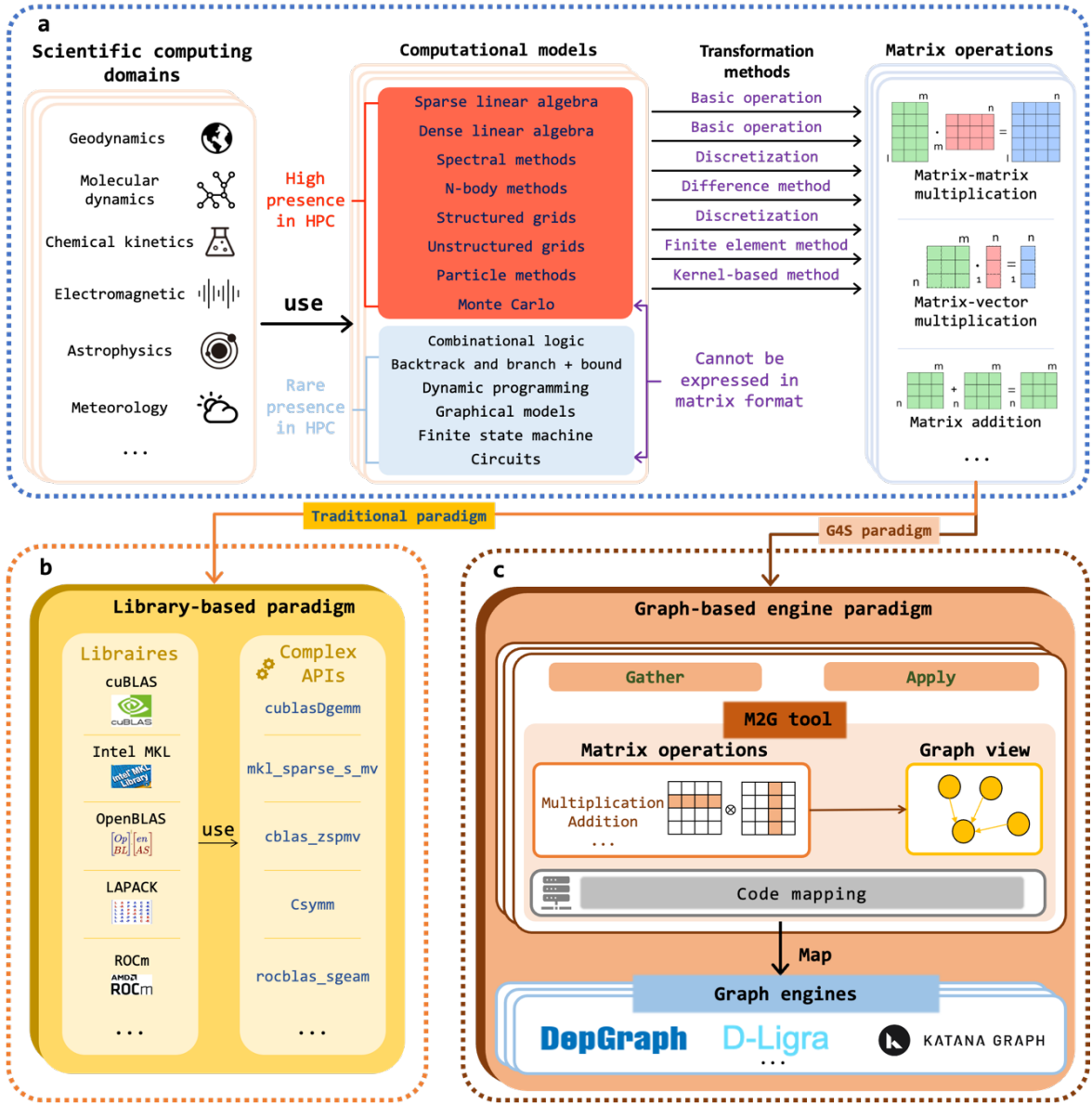

**Fig. 1. The comparison of the traditional programming paradigm and the G4S paradigm.** **a**, The computational models for scientific computing; the frequently used computational are marked with a red background, and those that are rarely used are marked with a blue background. **b**, Different libraries for scientific computing and various APIs they used. **c**, The graph-based engine paradigm, in which Gather and Apply are the only two programming interfaces that the users need to use and implement, the M2G tool transforms matrix operations into graph operations, and the code mapping component automatically determine the optimal strategies to run graph operations in the underlying graph engines.

To evaluate the performance of the proposed G4S paradigm, we have used it to implement a diverse set of representative matrix operations and the real-world scientific computing routines in three application domains: geodynamics [10], molecular dynamics [7], and chemical kinetics [11]. When running on multiple existing graph engines, these implementations achieve competitive performance in comparison with the cutting-edge library-based or bespoke implementations, which include CitcomS [18], DeePMD-kit [19], and

| Matrix operations \ Libraries | cuBLAS / cuSPARSE | Intel MKL | LAPACK | ROCm | OpenBLAS | G4S |
|---|---|---|---|---|---|---|
| Matrix-matrix addition | cublas<t>geam | mkl_<t>omatadd | - | rocblas_<t>geam | - | |
| Banded matrix-vector multiplication | cublas<t>gbmv | cblas<t>gbmv | <t>gbmv | rocblas_<t>gbmv | cblas<t>gbmv | |
| Matrix-vector multiplication | cublas<t>gemv | cblas<t>gemv | <t>gemv | rocblas_<t>gemv | cblas<t>gemv | |
| Symmetric banded matrix-vector multiplication | cublas<t>sbmv | cblas<t>sbmv | <t>sbmv | rocblas_<t>sbmv | cblas<t>sbmv | |
| Symmetric packed matrix-vector multiplication | cublas<t>spmv | cblas<t>spmv | <t>spmv | rocblas_<t>spmv | cblas<t>spmv | |
| Symmetric matrix-vector multiplication | cublas<t>symv | cblas<t>symv | <t>symv | rocblas_<t>symv | cblas<t>symv | |
| Packed symmetric rank-1 update | cublas<t>spr | cblas<t>spr | <t>spr | rocblas<t>spr | cblas<t>spr | |
| Packed symmetric rank-2 update | cublas<t>spr2 | cblas<t>spr2 | <t>spr2 | rocblas<t>spr2 | cblas<t>spr2 | |
| Symmetric rank-1 update | cublas<t>syr | cblas<t>syr | <t>syr | rocblas<t>syr | cblas<t>syr | |
| Symmetric rank-2 update | cublas<t>syr2 | cblas<t>syr2 | <t>syr2 | rocblas<t>syr2 | cblas<t>syr2 | |
| Triangular banded matrix-vector multiplication | cublas<t>tbmv | cblas<t>tbmv | <t>tbmv | rocblas<t>tbmv | cblas<t>tbmv | |
| Triangular banded linear system with a single RHS# | cublas<t>tbsv | cblas<t>tbsv | <t>tbsv | rocblas<t>tbsv | cblas<t>tbsv | |
| Triangular packed matrix-vector multiplication | cublas<t>tpmv | cblas<t>tpmv | <t>tpmv | rocblas_<t>tpmv | cblas<t>tpmv | |
| Packed triangular linear system with a single RHS# | cublas<t>tpsv | cblas<t>tpsv | <t>tpsv | rocblas<t>tpsv | cblas<t>tpsv | |
| Triangular matrix-vector multiplication | cublas<t>trmv | cblas<t>trmv | <t>trmv | rocblas_<t>trmv | cblas<t>trmv | Gather Apply |
| Triangular linear system with a single RHS# | cublas<t>trsv | cblas<t>trsv | <t>trsv | rocblas_<t>trsv | cblas<t>trsv | |
| Hermitian matrix-vector multiplication* | cublas<t>hemv | cblas<t>hemv | <t>hemv | rocblas_<t>hemv | cblas<t>hemv | |
| Hermitian rank-1 update* | cublas<t>her | cblas<t>her | <t>her | rocblas<t>her | cblas<t>her | |
| Hermitian rank-2 update* | cublas<t>her2 | cblas<t>her2 | <t>her2 | rocblas<t>her2 | cblas<t>her2 | |
| Hermitian banded matrix-vector multiplication* | cublas<t>hbmv | cblas<t>hbmv | <t>hbmv | rocblas_<t>hbmv | cblas<t>hbmv | |
| Packed Hermitian rank-1 update* | cublas<t>hpr | cblas<t>hpr | <t>hpr | rocblas<t>hpr | cblas<t>hpr | |
| Packed Hermitian rank-1 update* | cublas<t>hpr2 | cblas<t>hpr2 | <t>hpr2 | rocblas<t>hpr2 | cblas<t>hpr2 | |
| Hermitian packed matrix-vector multiplication* | cublas<t>hpmv | cblas<t>hpmv | <t>hpmv | rocblas<t>hpmv | cblas<t>hpmv | |
| Matrix-matrix multiplication | cublas<t>gemm | cblas<t>gemm | <t>gemm | rocblas_<t>gemm | cblas<t>gemm | |
| Symmetric matrix-matrix multiplication | cublas<t>symm | cblas<t>symm | <t>symm | rocblas_<t>symm | cblas<t>symm | |
| Symmetric rank-k update | cublas<t>syrk | cblas<t>syrk | <t>syrk | rocblas_<t>syrk | cblas<t>syrk | |
| Symmetric rank-2k update | cublas<t>syr2k | cblas<t>syr2k | <t>syr2k | rocblas_<t>syr2k | cblas<t>syr2k | |
| A variation of the symmetric rank-k update | cublas<t>syrkx | - | - | rocblas_<t>syrkx | - | |
| Triangular matrix-matrix multiplication | cublas<t>trmm | cblas<t>trmm | <t>trmm | rocblas_<t>trmm | cblas<t>trmm | |
| Triangular linear system with multiple RHS# | cublas<t>trsm | cblas<t>trsm | <t>trsm | rocblas_<t>trsm | cblas<t>trsm | |
| Hermitian matrix-matrix multiplication* | cublas<t>hemm | cblas<t>hemm | <t>hemm | rocblas_<t>hemm | cblas<t>hemm | |
| Hermitian rank-k update | cublas<t>herk | cblas<t>herk | <t>herk | rocblas<t>herk | cblas<t>herk | |
| Hermitian rank-2k update | cublas<t>her2k | cblas<t>her2k | <t>her2k | rocblas<t>her2k | cblas<t>her2k | |
| A variation of the Hermitian rank-k update | cublas<t>herkx | - | - | rocblas<t>herkx | - | |
| Sparse matrix-vector multiplication | cusparse<t>csrmv | mkl_sparse_<t>_mv | - | hipsparse<t>csrmv | - | |
| Sparse matrix-matrix multiplication | cusparse<t>csrmm | mkl_sparse_spmm | - | hipsparse<t>csrmm | - | |

<t>: 's' or 'S': Real single-precision; 'd' or 'D': Real double-precision; 'c' or 'C': Complex single-precision; 'z' or 'Z': Complex double-precision
#RHS: right-hand-side. * Only for complex numbers

**Fig. 2. Different types of matrix operations as well as the APIs used by different parallel computing libraries to implement these matrix operations.** In contrast, only two APIs (i.e., *Gather*() and *Apply*()) are needed to implement various matrix operations in G4S.

Cantera [20]. CitcomS is the best-performing routine with the bespoke implementation (i.e., not being implemented with the existing libraries in the market) for geodynamics. DeePMD-kit and Cantera are the well-optimized implementations for molecular dynamic and chemical kinetics, respectively. Both implementations are based on the state-of-the-art parallel computing libraries, such as MPI, TensorFlow, cuBLAS, and LAPACK.

## 2 MOTIVATION

### 2.1 The Graph View of Matrix Operations

Scientific computing routines from different domains typically involve various computational models [47], [48]. Fig. 1a shows a list of computational models in the literature. Among these models, the sparse/dense linear algebra [40], spectral methods [41], *N*-body methods [42], structured/unstructured grids [43], [44], particle methods [45], and Monte Carlo [46] are frequently used in high-performance scientific computing, while the others, including combinational logic, graphical models, dynamic programming, backtrack and branch+bound, finite state machine, and circuits, are rarely used [47], [48]. All frequently used computational models except for Monte Carlo can be transformed into matrix operations using the transformation methods listed in Fig. 1a (highlighted blue). Therefore, they are typically implemented as matrix operations in the state-of-the-art scientific computing libraries [18]-[20]. The remaining frequently used computational model, i.e., Monte Carlo, and other rarely used models, cannot be transformed into matrix operations.

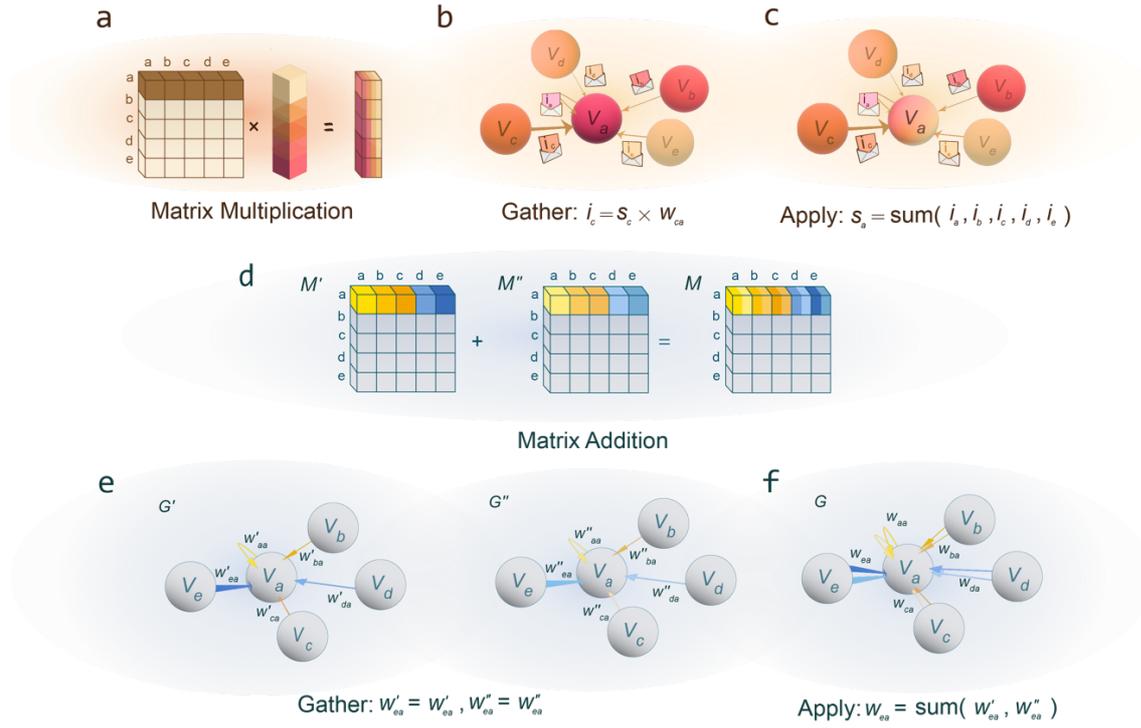

**Fig. 3. Illustration of transforming matrix operations to graph operations. a**, Matrix-vector multiplication (MV). **b**, The *Gather* operation for the MV: each vertex (e.g., $v_a$) gathers the information from its neighbors (e.g., $i_c$ from vertex $v_c$), where the information is the product of its neighbors' states (e.g., the state $s_c$ of the vertex $v_c$) and the weight of the edge between the neighbor and $v_a$ (e.g., the weight $w_{ca}$ of the edge $e_{ca}$). **c**, The *Apply* operation for the MV: it sums up the gathered information (i.e., $i_a$, $i_b$, $i_c$, $i_d$, and $i_e$) to update the state $s_a$ of $v_a$. **d**, Matrix addition, where the matrices $M'$ and $M''$ are added to obtain a new matrix $M$. **e**, The *Gather* operation for matrix addition: it collects the edge weights (e.g., $w_{ea}'$ and $w_{ea}''$) from two graphs (i.e., $G'$ and $G''$, which correspond to $M'$ and $M''$, respectively) for a vertex $v_a$. **f**, The *Apply* operation for matrix addition: it sums up the gathered edge weights (e.g., $w_{ea}'$ and $w_{ea}''$) to obtain the weight of the edge between $v_a$ and its neighbor in the graph $G$ ($G$ corresponds to the resulting matrix $M$).

## 2.2 Problems of the Traditional Solutions and Our Observations

Fig. 2 summarizes the types of matrix operations and the libraries we often see in scientific applications as well as the corresponding APIs the users have to invoke to implement the matrix operations with the libraries. Typically, when implementing a matrix operation using the library-based paradigm, the users have to select an appropriate library according to both the supporting computing devices and the features of input matrices, and then invoke the appropriate APIs provided the library. As listed in Fig. 2, the libraries do not provide a unified set of APIs to implement matrix operations. Specifically, when programming on NVIDIA GPUs, the users typically utilize the cuBLAS and cuSPARSE libraries, while on AMD GPUs, they often use the ROCm library. On the other hand, Intel MKL, LAPACK, and OpenBLAS libraries are widely used for CPU-based programming. In addition, to achieve an optimal performance, the type of matrix operation (e.g., addition, multiplication, rank-k update) and features of input matrices (e.g., sparse, dense, symmetric, triangular, packed, banded, Hermitian matrix) also affect the choice of APIs.

In contrast, with the GS4 paradigm, the users only need to use two unified programming interfaces: *Gather* and *Apply*. G4S views matrix operations from a graph perspective. In G4S, each non-zero element in a matrix is viewed as a data relationship (i.e., represented as an edge in a graph) between two data items (i.e., two vertices in a graph). Moreover, the actual value of a non-zero element is regarded as the weight of the edge. Specifically, for any matrix $A$ with $x$ rows and $y$ columns, matrix $A$ can be viewed as a graph with $m$ vertices, where the value of $m$ equals to the larger value of $x$ and $y$. If the cell $(i, j)$ in $A$ (denoted by $A[i,j]$) is a non-zero element, there is an edge $e_{ij}$ between vertices $v_i$ and $v_j$, and the value of $A[i,j]$ is the weight of edge $e_{ij}$ between vertices $v_i$ and $v_j$. By such means, any matrix can be transformed into a graph.

After converting a matrix into a graph, the matrix operations, e.g., matrix multiplication and matrix addition, can then be transformed into the graph processing operations. Take the matrix-vector multiplication as an example. The standard calculation is to multiply each element in each row of matrix $A$ by the corresponding element in the vector $V$, and then add up all products (Fig3.a). In G4S paradigm, for the $j$-th element in the vector $V$ (denoted by $V[j]$), its value can be viewed as the state $s_j$ of vertex $v_j$ in a graph. Since the value of $A[i,j]$ is the weight $w_{ij}$ of the edge $e_{ij}$, the multiplication of $A[i,j]$ by $V[j]$ can be viewed as the weight $w_{ij}$ of edge $e_{ij}$ multiplying by the state $s_j$ of vertex $v_j$ in graph processing. This operation is actually the *Gather* operation in graph processing, as illustrated in Fig. 3b. Furthermore, the multiplication of $A[i,j]$ by $V[j]$ needs to be summed up over j to obtain the multiplication of the $i$-th row of $A$ by the vector $V$. The summation operation is equivalent to the *Apply* operation in graph processing, i.e., performing the accumulation over the state of every neighbor of vertex $v_i$ multiplying the weight of the corresponding edge between vertex $v_i$ and the neighbor. This is illustrated in Fig. 3c.

As for matrix addition, e.g., $M_3 = M_1 + M_2$ (illustrated in Fig. 3d), the addition of the corresponding elements in matrices $M_1$ and $M_2$ can be viewed as merging the edges (i.e., adding their edge weights) of the graphs corresponding to $M_1$ and $M_2$. Here, taking the values of the individual edge weights is the *Gather* operation, while adding their edge weights is the *Apply* operation. This transformation is illustrated in Fig. 3e,f.

Different from the traditional scientific computing programming paradigm, which treats the non-zero elements in the matrix as a set of discrete data, the G4S paradigm views the non-zero elements in the matrix as the weights of the edges between vertices in a graph. The graph view of a matrix clearly depicts the data relationship between elements by utilizing graph edges, which opens up many new opportunities, such as load balance, communication optimization, for the G4S paradigm to optimize performance. More importantly, through the G4S paradigm, the users only need to invoke two unified graph processing programming interfaces: *Gather*() and *Apply*(), which significantly simplifies the programming of scientific computing routines.

```
INPUT: stiffness matrix A, velocity vector u.
OUTPUT: The boundary forces f.

 1: function Gather(A, u)
 2:    for each edge $e_i$ of the vertex v do
 3:       $ft_i \leftarrow v.u_i \cdot e_{ij}.weight$   // u: velocity of the mantle point
                                      // weight: stiffness between two neighboring mantle points
                                      // $ft_i$: a temporal result for v
 4:    end for
 5:    return ft                // ft: the gathered results (i.e., a set of $ft_i$) for v
 6: end function

 7: function Apply(u, ft)
 8:    for each $ft_i$ of vertex v do
 9:       $v.f \leftarrow v.f + ft_i$
10:    end for
11:    return v.f               // v.f: the boundary force for vertex v
12: end function
```

**Fig. 4. The implementation of the matrix computations from geodynamics (i.e., CitcomS) using the unified APIs, Gather() and Apply(), provided by the G4S paradigm.** The M2G tool automatically converts the matrix computations into graph operations. The user only needs to 1) customize the *Gather*() to collect the results of the velocity of each mantle point (e.g., vertex *v*'s state in the graph) multiplying by the stiffness between this mantle point and its neighboring mantle points (i.e., the weight of *v*'s edge $e_i$), and 2) program *Apply*() to accumulate the gathered results to obtain the boundary forces.

## 3     THE IMPLEMENTATION OF SCIENTIFIC COMPUTING ROUTINES BASED ON THE G4S PARADIGM

The G4S paradigm uses two interfaces, *Gather*() and *Apply*(), to represent all matrix operations in scientific computing routines. *Gather*() is used for each vertex to gather the information from its neighbors, while *Apply*() processes the gathered information to update the vertex state. In G4S, users are relieved from the burden of dealing with complex characteristics of input matrices and the optimization strategy for matrix operations. They only need to focus on implementing these two interfaces according to the specific purpose of matrix operations in the scientific computing routines. This allows users to carry out their tasks without requiring advanced HPC expertise.

To demonstrate the effectiveness and generality of G4S, we implemented a diverse set of matrix computations (i.e., those listed in Fig. 2) using G4S. Furthermore, we used G4S to implement the scientific computing routines in three real-world scientific applications, CitcomS, DeePMD-kit and Cantera, which are from the domains of geodynamics, molecular dynamics, and chemical kinetics.

CitcomS [18] uses finite element codes, which contain a large number of bespoke implementations of matrix-vector multiplications, for solving mantle convection problems in the field of geodynamics. The operation of mantle force calculation is a fundamental part of this application. Fig. 4 shows how to program mantle force calculation in CitcomS using the G4S paradigm, the user only needs to customize *Gather*() to collect the velocity of each mantle point (e.g., the state of a vertex in the corresponding graph) and multiply it by the stiffness between this mantle point and its neighboring mantle points (i.e., the weight of the edge between this mantle point and the neighboring point). In *Apply*(), the gathered results are accumulated to obtain the boundary forces.

DeePMD-kit [19] is the state-of-the-art implementation to model the interatomic potential energy and force field in molecular dynamics, where the core operation is the potential energy calculation, which includes a large number of matrix-matrix multiplications. When using G4S to represent this operation, the user only needs to i) initialize *Gather*() to multiply the results of the relative position of each atom (the state of a vertex in the graph) by the distance information between this atom and its neighboring atoms (i.e., the edge weight), and ii) implement *Apply*() to sum up the multiplication results obtained for each of vertex's neighbor to obtain the potential energy.

Cantera [20] is used to realize coal combustion to simulate chemical kinetics. The main operation of this routine is matrix-vector multiplication. We can use the G4S paradigm to represent the core operation of the gas combustion (i.e., the heat capacity calculation in a high-pressure reflected shock tube reactor). Specifically, the user only needs to i) initialize *Gather*() to find the results of the gas pressure of each chemical component (e.g., a vertex $v$'s state in the graph) and multiply them by the gas temperature of its neighbor chemical components (i.e., the weight of the edge between $v$ and its neighbor), and ii) utilizes *Apply*() to aggregate the calculated results in order to simulate the heat capacity after a certain time.

## 4 G4S METHODOLOGIES

### 4.1 Matrix-to-Graph Transformation and Code Mapping in G4S

In the G4S paradigm, a tool called M2G is developed. The M2G tool has two main functionalities. Firstly, it implements the matrix-to-graph transformation as discussed above (see Fig. 4). Secondly, it includes a code mapping mechanism to map the transformed graph operations to run on the underlying graph engine, i.e., use the optimal graph engine to implement different kinds of matrix operations. The core component of the code mapping mechanism is a decision tree. It automatically (transparent from the users) classifies the graph operations and determines the optimal strategies, including the strategies for graph preprocessing, graph execution, and communication, for processing the graphs on the underlying graph engine based on the data relationships exposed in the graphs. To train the decision tree, we label different matrix operation types (e.g., addition or multiplication), different characteristics of input matrices (e.g., sparse, dense or triangular) and different hardware platforms (e.g., CPU, GPU, or hybrid platform). The decision tree is trained based on the above labels and the ground-truth optimal graph processing strategies. The trained decision tree is then used to determine the optimal graph processing for the input matrix operations. Afterwards, the equivalent graph operations are dispatched to the underlying graph engine and are run adaptively with the optimal strategies output by the decision tree. The code mapping mechanism allows matrix operations to be run on a wide variety of large-scale computational platforms without the intervention of HPC experts.

## 4.2 Design of the M2G Tool

We develop a tool called M2G for automatic transformation of matrices into graphs. It first identifies the matrix data from the input datasets by checking if each row in the input dataset has the same number of elements. Second, matrix data should usually be of numeric type (e.g., integer and floating-point), M2G determines the type in the matrix data (which are usually of numeric type, e.g., integer or real number). Then, each non-zero element in the input matrix is treated as an edge connecting two vertices in the resulting graph, which captures the relationships between physical representations of rows and columns in the matrix and establishes the dependencies among matrix elements (e.g., there exist dependency among the vertices and edges on the same directed path in the graph).

Additionally, M2G automatically transforms matrix operations, which are implemented through the two Gather and Apply interfaces in G4S, into graph operations. Transformation of matrix-vector multiplication and matrix addition are discussed in the main article. In addition, for matrix-matrix multiplication, it can be considered as a series of matrix-vector multiplication. Specifically, for a matrix $B$ with $a$ rows and $b$ columns and a matrix $C$ with $c$ rows and $d$ columns, $B$ multiplying by $C$ can be viewed as merging $d$ different matrix-vector multiplications. Through this transformation, the capabilities of a graph engine, which is able to handle complex graph processing tasks effectively, can fully exploited. M2G also includes a caching mechanism to enhance performance. Since matrices are often processed more than once in scientific computing routines, M2G automatically caches the graphs transformed from the matrices. The cached graphs will be reused whenever possible, which amortizes the transformation overhead.

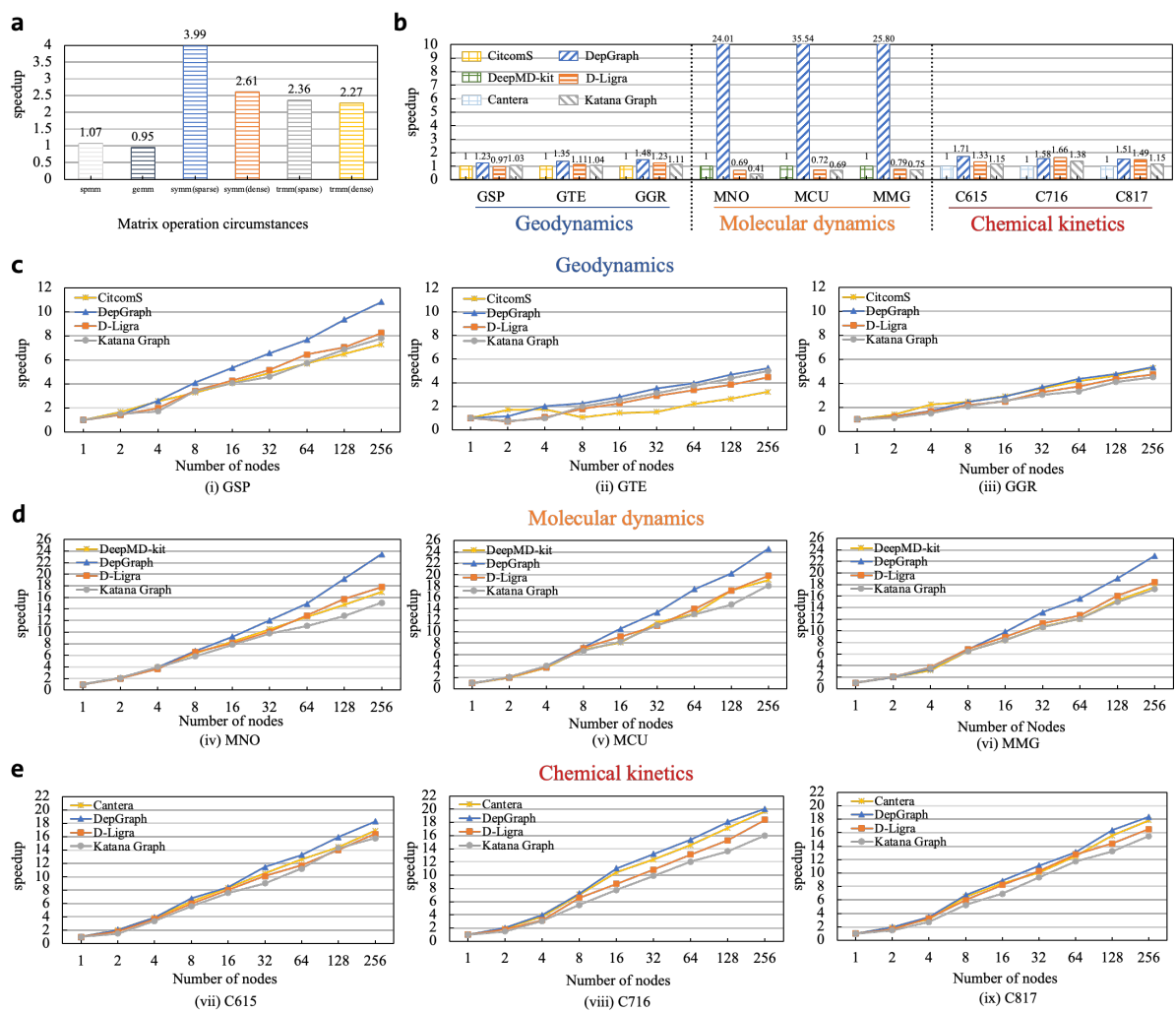

**Fig. 5. The performance of the scientific computing routines based on the G4S paradigm in comparison with the state-of-the-art scientific computing implementations**. **a**, The speedup of matrix operations in different scenarios. **b**, The speedup of the scientific computing routines in geodynamics, molecular dynamics, and chemical kinetics domains programmed, being implemented by the G4S paradigm and executed by three graph engines (i.e., DepGraph, D-Ligra, and Katana Graph) on a distributed computing platform, in comparison with the speedup of the state-of-the-art implementations on the same platform. **c,d,e**, The scalability of different implementations on the distributed platform. The distributed platform consists of 256 machines interconnected by a 400 Gbps InfiniBand network. Each machine has two 64-core 3.0 GHz Kunpeng 920 CPUs and 256 GB main memory.

## 4.3 Graph-Based Optimization Methods for Scientific Computing

Fig. 5 shows that graph-based techniques achieve comparable performance to the state-of-the-art implementations for different sets of matrix operations. In this section, we introduce the graph-based optimization methods used to support scientific computing from the following three aspects.

**Graph-Based Processing.** The traditional scientific computing approaches mainly rely on matrix reordering to reduce the processing of zero elements. However, the resulting irregular data accessing pattern leads to poor data locality. In contrast, when transforming a matrix with irregular sparsity to a graph, zero elements are eliminated. Moreover, the matrix elements that are likely to manifest high locality will naturally become closely connected in the converted graph. A set of closely connected vertices form a vertex community. Next, the traversal strategy traverses the vertices in each community consecutively to maximize the neighbor sharing among vertices (i.e., reduce the number of times a vertex has to be accessed repeatedly). After that, the graph is reordered by assigning the vertices in a community with continuous IDs along the traversing order. Note that the above reordering process is lightweight in both computation and memory cost, and can be parallelized effectively.

To effectively parallelize the processing of the transformed graph over a distributed platform, the vertices and their edges of the graph are evenly partitioned among different machines (a partition of vertices and their edges is called a subgraph). By such means, compared with the traditional scientific computing approaches, the closely-connected vertices will be dispatched to the same machines, and be more likely to be processed with better temporal and spatial locality.

**Graph-Based Execution Model.** G4S paradigm enables to achieve fine-grained data parallelism to achieve higher degree of parallelism and smaller synchronization cost. Specifically, after the transformed graph has been partitioned among different machines, the graph-based techniques allow each machine to efficiently handle its corresponding subgraph in a fine-grained data parallelism mechanism, e.g., taking each vertex or edge as the parallel processing unit. To solve the problem of the high irregularity of the matrices (which incurs irregular workload and irregular data access), the fine-grained data parallelism mechanism also allows to transform the irregular data into more regular ones according to the data relationships potentially shown in the graph. In detail, it identifies vertices with high degrees (i.e., more other vertices depend on these vertices) and then splits these vertices until their degrees reach a predefined limit (the value is ten by default). Next, the split vertices can be adaptively dispatched to multiple cores along the data relationships for efficient parallel processing and balanced load.

To achieve high degree of data parallelism for the sequential multiplication of a series of matrices, G4S paradigm decouples the data dependencies between the non-zero elements in multiple matrices. In detail, it constructs direct dependencies for the non-zero elements along data dependency and then enables these non-zero elements to be handled concurrently using these direct dependencies, where a long data dependency between two non-zero elements along a series of matrices is represented as a direct dependency between these two non-zero elements. Note that, when using existing solutions, these non-zero elements have to be sequentially handled along the data dependency.

Besides, to efficiently utilize the main memory bandwidth via alleviating the irregular data access, the graph-based techniques can fuse the updates of the vertices to improve the spatial locality, where the updates of the vertices with consecutive IDs (i.e., a bucket) are batched and applied together. To further reduce synchronization cost, each bucket is typically assigned to be handled by one core. To dynamically ensure the balanced load among different machines, some graph data of the overloaded machine are migrated to the machine with the lower load when the spared execution time is evaluated to be more than the migration time. In this way, for the processing of different types of matrices in scientific computing routines, the G4S paradigm allows to effectively reduce data access and synchronization costs, and ensure load balance.

**Graph-Based Communication Scheme.** The communication schemes adopted in the traditional scientific computing methods cannot perceive data dependencies, incurring significant irregular communications and data synchronization overheads on a distributed platform. In comparison, a graph engine allows to adaptively regularize the communications according to the data relationships, efficiently reducing the communication and data synchronization overhead on a distributed platform.

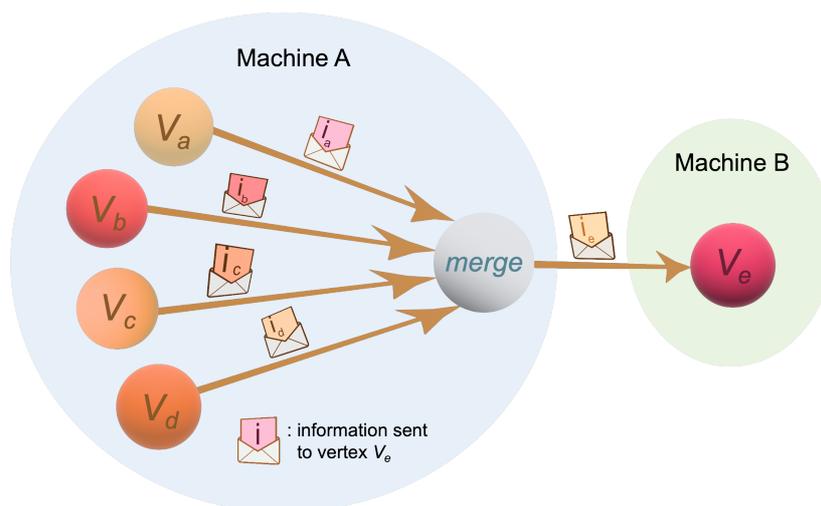

**Fig. 6. Communication merge phase.** Before the information (i.e., the state of each vertex and the weight of each edge) is sent from machine A to the designated machine B, all the information will be merged together, thus realizing a single piece of information from lots of information, which greatly reduces the communication and caching overhead.

Specifically, when a machine has completed the processing of a subgraph assigned to it, it typically needs to synchronize its local results by communicating the results of its subgraph to other machines. To regularize the communications, a graph engine can batch the communications (see Fig. 6) in the buffer until the information about individual data to be communicated is fully merged. The information regarding which communications should be batched can be obtained according to the data relationships exhibited by the graph structure. After batching the communications in each machine, the communications across different machines are further grouped according to their destination machines. The communications destined to the same machine will be sent simultaneously to alleviate the irregular communications, improving utilization of the network bandwidth. To reduce the communication volume, the information can be encoded to avoid the communications of the unnecessary information (e.g., vertices IDs). To further hide the communication overhead, asynchronous communication strategies can be employed to overlap computation and communication.

Moreover, when processing the graph converted from the operation on matrices with irregular sparsity, the graph engine can automatically identify the characteristics of the data (e.g., high degree vertices) from the data relationships and employ different communication strategies accordingly. For example, to optimize the communications associated with high degree vertices, the replications of these data will be created on all machines, because their connected data are typically existed on almost every machine. The other vertices (except for the high degree vertices), on the other hand, will remain connected to other data without replications because they have minimal connections and delegating them would offer little benefit while requiring additional time and space to manage the replications.

### 4.3 Details of Code Mapping of G4S Paradigm

To efficiently perform the transformed graph operations on the underlying graph engine, we design a code mapping mechanism. It first explores the optimal execution strategy for a graph operation on the graph engine (e.g., DepGraph [13], D-Ligra [14], and Katana Graph [15]), according to the data relationships exposed in the graph. The graph operation is then dispatched accordingly to the underlying graph engine for execution.

To explore the optimal execution strategy, a decision tree is developed according to the performance of different execution strategies performed on different types of matrix operations, different characteristics of input matrices, and different hardware platforms. Different matrix operations are automatically classified by this decision tree to explore the optimal optimization strategies.

Specifically, different matrix operations (consequently different graph operations) need different parallelization execution strategies (e.g., the vertex-centric and edge-centric parallel execution model for matrix multiplication and addition, respectively) to achieve optimal performance. Note that the memory access pattern is different for matrix multiplication and addition, where the former performs many irregular memory accesses while the memory accesses in the latter is regular.

To efficiently support the processing of different types of input matrices (e.g., dense, sparse, symmetric, triangular, packed, banded, and Hermitian matrices, which correspond to different characteristics of the graphs, different choices should be made for the execution strategy. For example, when processing dense matrices (i.e., the vertices in the converted graph are closely connected [16]), it can apply the fine-grained data parallelism mechanism to achieve a high degree of parallelism and low synchronization cost. However, when processing the matrices with irregular sparsity, it needs to preprocess the matrices appropriately to achieve better data locality, transform irregular data into more regular ones, and also apply differentiated communications [52]-[54] and scheduling strategies to achieve optimal performance.

When handling the transformed graph operations on different hardware platforms (e.g., CPU, GPU, or hybrid), it needs to employ different optimization strategies too. For example, to obtain high performance on GPU-based platform, it needs to ensure coalescing memory accesses to utilize high memory bandwidth, exploit multilevel (thread-level, warp-level, and CTA-level) parallelism for high execution efficiency, and employ the asynchronous execution scheme to overlap communication and computation. Nevertheless, for the hybrid platform, it can utilize the graph partitioning scheme to dispatch the graph processing tasks to different devices (e.g., CPU and GPU) (e.g., CPUs are better at handling irregular tasks GPUs) and employ efficient communication strategies to minimize data transfers between different types of devices to reduce communication overhead.

When the optimal execution strategy has been obtained, the graph operation is dispatched to be efficiently executed on the underlying graph engine according to the strategy. Specifically, the M2G tool needs to know the programming models and APIs of existing graph engines. For example, to implement the graph algorithms on DepGraph, it needs to instantiate the functions of *Accum*() and *EdgeCompute*(), respectively, where the *Accum*() is used for each vertex to gather the states of its neighbors and the *EdgeCompute*() is used for each vertex to calculate its new state and scatter this state to its neighbors. Meanwhile, according to the transformed graph operations and the obtained optimal execution strategy, it creates an abstract intermediate representation of the graph operation that separates the algorithmic logic from the specific details of each graph engine. After that, it develops the specific implementations of the graph operation for each graph engine by leveraging the programming model and APIs of this engine to translate the abstract intermediate representation into engine-specific codes. Note that the code mapping mechanism can automatically dispatch the graph operation to several well-known graph engines (e.g., DepGraph, D-Ligra, and Katana Graph), and we will develop automatic code mapping for more graph engines in the future.

**Table 1. The datasets in the real-world scientific computing routines**

| Dataset | Description |
|---|---|
| GD_speed (GSP) | Thermal convection simulation dataset with velocity boundary conditions imposed on the top surface in a given region of the sphere in geodynamics. |
| GD_temp (GTE) | Natural convection simulation dataset in the presence of variable viscosity, including temperature-dependent or stress-dependent viscosity in geodynamics. |
| GD_grid (GGR) | Natural convection simulation dataset in the presence of variable viscosity explored with a fine grid in geodynamics. |
| MD_water (MWA) | Molecular structure of water in molecular dynamics. |
| MD_cuprum (MCU) | Molecular structure of cuprum in molecular dynamics. |
| MD_fparam (MFP) | Molecular structure of oxygen with electron temperature based on temperature dependent deep potential in molecular dynamics. |
| CK_3072 (C3072) | Ignition delay time computations in a high-pressure reflected shock tube reactor based on a stoichiometric mixture of 3072 kinds of gas. |
| CK_4096 (C4096) | Ignition delay time computations in a high-pressure reflected shock tube reactor based on a stoichiometric mixture of 4096 kinds of gas. |
| CK_5120 (C5120) | Ignition delay time computations in a high-pressure reflected shock tube reactor based on a stoichiometric mixture of 5120 kinds of gas. |

## 5. EXPERIMENTAL RESULTS

To evaluate the performance of the G4S paradigm, we first use micro-benchmarks to measure the performance of a diverse set of matrix calculations, including sparse matrix-matrix multiplication (spmm), dense matrix-matrix multiplication (gemm), symmetric matrix-matrix multiplication (symm), and triangular matrix-matrix multiplication (trmm). In addition to microbenmarking, we use the interfaces provided by G4S to implement the matrix operations in three typical scientific computing routines from real-world scientific applications: CitcomS [18], DeePMD-kit [19], and Cantera [20], which perform different types of matrix operations and adopt different implementations as discussed in previous sections. The G4S-based implementations are executed on three popular graph engines: DepGraph [13], D-Ligra [14], and Katana Graph [15]. The performance of the implementations is compared with the current state-of-the-art implementations of these three applications. The datasets used in the comparison are shown in Table 1.

Figure 5a,b shows that the G4S-based implementations achieve comparable performance to the state-of-the-art implementations for these different sets of matrix operations as well as for CitcomS, DeePMD-kit, and Cantera. Notably, these traditional implementations, whether library-based or custom, have undergone continuous optimizations for many years, if not decades. Despite this, G4S showcases its effectiveness in achieving competitive performance levels.

In some cases, G4S even improves the performance significantly. For example, for DeePMD-kit, DepGraph [13] achieves up to 32.62 times speedup in comparison with its state-of-the-art implementation [1] that won the 2022 ACM Gordon Bell Prize finalist, and up to 240.08 times speedup compared with the results [2] presented in the winner of 2020 ACM Gordon Bell Prize. It is because DepGraph can efficiently decouple the complex data dependencies [49] between the non-zero elements in multiple matrices when multiplying a series of matrices together in DeePMD-kit, ensuring higher degree of data parallelism. Besides, as shown in Fig. 5c-e, the G4S-based implementations achieve the competitive scalability compared with the state-of-the-art traditional implementations.

# 6 RELATED WORK

## 6.1 High-Performance Computing Techniques for Scientific Computing Routines

Several studies have focused on optimizing matrix operations in the context of high-performance scientific computing. These research efforts aim to enhance the performance and efficiency of matrix-based algorithms commonly used in scientific simulations and analyses.

One important area of study [55] looked at effective matrix operation parallelization methods. Previous research [56] has suggested new parallel matrix multiplication algorithms that successfully use distributed memory architectures and significantly increase performance on large-scale clusters.

Another area of study has concentrated on improving matrix factorizations. Previous studies [57] looked into how different factorization methods, including LU and QR factorization, performed on modern multi-core processors. For large-scale singular value decomposition, a hybrid strategy combining distributed computing and GPU acceleration was suggested in a related study [58]. This method demonstrated substantial performance increases by efficiently leveraging both CPU and GPU resources.

Additionally, memory access patterns in matrix calculations have been optimized through study. Cache-aware matrix transposition algorithms have been developed in the past with the aim of reducing cache misses and enhancing memory locality [59]. These methods have proven to perform better than traditional transpose algorithms, especially for big matrices. Additionally, sparse matrix-vector multiplication performance on modern architectures has been greatly enhanced by the introduction of memory layout optimization techniques for sparse matrices [60]. These techniques enable effective access patterns and lower storage requirements.

## 6.2 Graph Processing Techniques

Numerous studies have looked into how to perform matrix operations using graph processing methods, offering fresh ideas for improving the speed and scalability of matrix operations.

Distributed matrix operation frameworks based on graphs have attracted a lot of interest. Iterative graph algorithms can be executed effectively thanks to the widely used framework GraphX [61], which combines distributed linear algebra and graph processing. Another well-known framework, GraphLab [62], uses a vertex-centric programming model to facilitate parallel computation of graphs, enabling effective matrix operations in distributed systems. GraphLab is more efficient and scalable than traditional synchronous models because it permits vertices to update their states asynchronously and propagate information throughout the graph. Pregel [63] provides an abstraction for graph computations and focuses on scalability and fault tolerance in distributed environments. Ligra [14] is a lightweight framework which simplifies the graph traversal algorithms.

METIS [64] developed a multilevel approach that effectively divides unbalanced graphs into balanced subgraphs while reducing communication between partitions. GAP [65] takes a deep learning approach to perform graph partitioning, which defines a differentiable loss function to represent the partitioning objective.

PowerGraph [66] introduces the vertex-centric Gather-Apply-Scatter programming model, which has become a defining feature of graph processing frameworks, and offers a potent abstraction for graph algorithms that makes processing large-scale graphs effective and scalable. However, it has only been shown to be effective in natural graphs (e.g., social network graphs, web graphs), and it remains unknown whether it is effective in graphs for high-performance computing. GraphMat [67] uses vertex-centric programming on the front-end and maps it to high-performance sparse matrix operations on the back-end which achieve better adjacency matrix partitioning and improved load balancing between threads.

To optimize particular matrix operations, graph-based algorithms have been created in addition to frameworks and partitioning strategies. In order to facilitate efficient and scalable matrix multiplication on a range of architectures, GraphBLAS [68] provides a high-level abstraction for matrix operations on graphs. The GraphBLAST package [69], which offers high-performance implementations of GraphBLAS operations on CPUs and GPUs, is one of the most notable examples. Efforts [70] are also being made to optimize and parallelize its operations for new hardware architectures like FPGA and specialized graph processing units.

## 7 DISCUSSION

G4S introduces a transformative programming paradigm shift in high-performance scientific computing. In the existing library-based programming paradigm, it is essential to select an appropriate library, often with its own set of APIs, to implement the computations. Additionally, careful attention has to be paid to design the strategies of data distribution and computing parallelization, which plays a crucial role in achieving high performance on a large-scale platform. However, all these necessitate extensive involvement of HPC experts. Besides, the traditional library-based paradigm predominantly employs the coarse-grained task parallelism, which limits the ability to fully exploit dependencies between data and harness the potential of complex data parallelism in scientific computing routines, resulting in suboptimal performance. In contrast, when using G4S to program scientific computations, users only need to work with two graph processing APIs: *Gather*() and *Apply*(), leaving all the optimization decisions to the underlying graph engine. This frees HPC experts from the programming of scientific computations. Moreover, since the data relations are embedded in the graph after the matrix-to-graph transformation, it lends many optimization opportunities to the graph engine, which is the fundamental reason why the G4S-based implementations can achieve competitive performance compared with the existing state-of-the-art implementations.

In the future, we will study an even wider range of matrix operations and more approaches to optimize the execution of scientific computing routines on existing graph engines. Moreover, we will develop a high-performance and fault-tolerant graph engine to support large-scale scientific computing routines. Finally, we plan to investigate the feasibility of using graph engines for the operations that cannot be represented as the matrix computations, such as Monte Carlo, which will enable a broader range of scientific computing routines to be implemented using the graph-based techniques.